\documentstyle[aps,preprint,tighten,amsfonts]{revtex}

\newtheorem{theorem}{Theorem}

\begin{document}

\title{BOSE-EINSTEIN CONDENSATION IN THE PRESENCE
OF A UNIFORM FIELD AND A POINT-LIKE IMPURITY}
\author{R.\ M.\ Cavalcanti\footnote{E-mail: rmoritz@if.ufrj.br}}
\address{Instituto de F\'{\i}sica, Universidade Federal do
Rio de Janeiro \\
Caixa Postal 68528, 21945-970 Rio de Janeiro, RJ, Brazil}
\author{P. Giacconi,\footnote{E-mail: Paola.Giacconi@bo.infn.it} 
G. Pupillo,\footnote{Present address: Laboratory for Physical Sciences,
8050 Greenmead Drive College Park, MD 20740; E-mail: 
Guido.Pupillo@physics.umd.edu} 
and R. Soldati\footnote{E-mail: Roberto.Soldati@bo.infn.it}}
\address{Dipartimento di Fisica, Universit\'a
di Bologna and Istituto Nazionale \\
di Fisica Nucleare, Sezione di Bologna, 40126 Bologna, Italia}   
\maketitle
                    
\begin{abstract} 

The behavior of an ideal $D$-dimensional boson gas in the presence of a 
uniform gravitational field is analyzed. 
It is explicitly shown that, contrarily to an old standing folklore,
the three-dimensional gas does not undergo Bose-Einstein condensation
at finite temperature.
On the other hand, Bose-Einstein condensation occurs at $T\neq 0$
for $D=1,2,3$ if there is a point-like impurity at the bottom of the
vessel containing the gas.

\end{abstract}


\vspace{1cm}
\centerline{{\sl Accepted for publication in Physical Review A}}
\vfill
\noindent
DFUB/14/01
\hfill
November 2001
\eject


\section{Introduction}

The response of quantum systems to the influence of external background
fields is of utmost importance in a wide number of physical applications.
As well, the role of disorder, i.e., the presence of impurities in 
condensed matter systems, is often crucial in the occurrence of remarkable
physical effects. It is the aim of the present paper to investigate the
behavior of an ideal boson gas in the presence of a uniform (i.e., constant
and homogeneous) gravitational field and of extremely localized 
(actually point-like) impurities affecting the quantum dynamics of
the bosonic particles.

It is well known since a long time \cite{Hua,Pat} that an ideal 
three-dimensional boson gas in free space 
undergoes a phase transition called 
{\em Bose-Einstein condensation} 
(BEC), in which a finite fraction of its constituent molecules 
condenses in the single-particle ground state. Such a condensation 
differs from the usual condensation of a vapor into a liquid in 
that there is no phase separation. For this reason, BEC is commonly 
described as a phase transition in momentum space --- the particles 
condense into the $|{\bf p}={\bf 0}\rangle$ state, which has a 
uniform spatial distribution. It is also well known \cite{Pat} that such a 
phase transition is no longer possible, for free bosons, 
in one and two dimensions --- although in both cases
it does occur in the presence of a point-like attractive 
potential \cite{IGH,GMS2}. A long standing popular belief 
\cite{Hua,Gol,Lam,Groot,Halpern,Gersch,Bagnato1}
is that if the particles of a 3D ideal boson gas were
placed in a (uniform) gravitational field, then BEC would still occur,
but in the condensation region there would be a spatial separation
of the two phases, just as in a gas-liquid condensation. 
 
In the present paper we study the exactly solvable quantum
mechanical model of an ideal boson gas in $D=1,2,3$ dimensions
in the presence of a uniform gravitational field and of a point-like
impurity formally described by a $\delta$-function potential.
In order to make the Hamiltonian bounded from below, so that
the system may attain a state of thermodynamic equilibrium, 
we shall enclose the gas in a container with impenetrable walls.
Concerning the mathematical description of a point-like
impurity, it should be remarked that a $\delta$-potential
is generally ill-defined when $D>1$, and
some renormalization procedure is mandatory.
Actually, the rigorous mathematical procedure to deal with point-like 
interactions involves the analysis of the self-adjoint extensions of the
symmetric Hamiltonian operator \cite{Alb}. 
In the present work, however, we prefer to
follow a more informal approach \cite{Jac} which is closer to the physical 
intuition, but reaches the same final result as the 
rigorous though more involved method of 
self-adjoint extensions \cite{ReS}.
To be specific, we formally treat the contact interaction as a 
$D$-dimensional 
$\delta$-potential, then proceed to the renormalization procedure
in physical terms, and finally obtain the so-called
Krein's formula for the Green's function, from which it is
possible to extract the energy spectrum of the single-particle
Hamiltonian. 

In Section \ref{no-imp} we prove that an ideal boson gas in the presence of
a uniform gravitational field does not undergo BEC at finite temperature, 
except in the one-dimensional case. This implies, in particular, that in the
three-dimensional case no phase separation occurs in the thermodynamic 
limit, at variance with the above quoted conventional wisdom.
We also provide a rather general {\em sufficient} condition for
the occurrence of BEC in a trapped ideal gas, which generalizes
some results obtained by other 
authors \cite{Bagnato2,Li,Yan,Yan2,Salasnich} for power-law potentials. 
In Section \ref{Floor} we show that the onset of BEC in a uniform 
gravitational field is made possible in $D=2,3$ if a
point-like impurity (i.e., a $\delta$-potential) is placed 
at the bottom of the vessel containing the gas.
The reason is that the presence of the 
impurity entails the existence of a bound state,
whose energy gap with respect to the continuous spectrum
is what is needed for the ideal gas to undergo BEC.
In Section \ref{Conclusions} we draw our conclusions, 
whereas some technical details are presented in two Appendices.
  

\section{$D$-dimensional boson gas in a uniform field}
\label{no-imp}

It is convenient to first analyze
and discuss the impurity-free case, which turns out to 
exhibit, as we shall see
below, rather surprising features. Thus,
in this Section we shall study the quantum mechanical behavior 
of an ideal boson gas in the presence of a uniform gravitational field.
The existence of a (single-particle) ground state is guaranteed by the presence
of an impenetrable wall at the bottom of the vessel containing
the gas. The single-particle Hamiltonian is given by
\begin{equation}
\label{4.1} 
H_0^{(D)}(g)={{\bf p}^2\over 2m}+mgx,
\end{equation}
in which we have set
\begin{equation}
{\bf x}=(x_1,\ldots,x_D)\equiv ({\bf r},x),\qquad
{\bf p}=(p_1,\ldots,p_D)\equiv ({\bf k},p).
\end{equation}
The gas is supposed to be enclosed in a rectangular
box of sides $L_1,L_2,\ldots,L_D$, with its bottom 
fixed at the plane $x=0$.
Since we are interested in the thermodynamic limit, we can,
without lack of generality, impose
periodic boundary conditions in the $x_1,\ldots,x_{D-1}$ directions
and Neumann boundary condition\footnote{The reason why 
we impose Neumann boundary condition,
instead of the seemingly more natural Dirichlet one,
will be explained in Section \ref{Floor}.}
at $x=0$ and $x=L_D$, i.e.,
\begin{equation}
\psi(x_1,\ldots,x_j+L_j,\ldots,x_D)=\psi(x_1,\ldots,x_j,\ldots,x_D),
\qquad j=1,\ldots,D-1,
\end{equation}
\begin{equation}
\partial_x\psi({\bf r},x=0)=\partial_x\psi({\bf r},x=L_D)=0,
\end{equation}
and then take the limits $L_j\to\infty$, $j=1,\ldots,D$.
After these limits are taken,
the eigenfunctions and eigenvalues of $H_0^{(D)}(g)$ read
\begin{equation}
\psi_{n,{\bf k}}({\bf r})=
{\exp\left\{(i/\hbar)\,{\bf k}\cdot{\bf r}
\right\} \over (2\pi\hbar)^{(D-1)/2}}\,
\sqrt{-\frac{\kappa}{a_n'}}\,
{{\rm Ai}(\kappa x+a_n')\over{\rm Ai}(a_n')}, 
\label{4.2} 
\end{equation}
\begin{equation}
E_{n,{\bf k}}={{\bf k}^2\over 2m}-E_g a_n'\,,\qquad
n\in{\mathbb N},\,\,{\bf k}\in{\mathbb R}^{D-1}, 
\label{4.3} 
\end{equation}
where ${\rm Ai}(x)$ is the Airy function \cite{AbS},
$a_n'$ are the zeros of ${\rm Ai}'(x)$, and the
parameters $\kappa$ and $E_g$ are defined as
\begin{equation}
\kappa\equiv\left({2m^2 g \over \hbar^2}\right)^{1/3},\qquad
E_g\equiv {mg\over \kappa}={\hbar^2\kappa^2\over 2m}.
\label{2.3}
\end{equation}
All the zeros of ${\rm Ai}'(x)$ are negative, 
hence the energy levels $E_{n,{\bf k}}$ are positive.

If $D>1$ the spectrum is purely continuous and the 
corresponding improper eigenfunctions 
are normalized according to
\begin{equation}
\langle\psi_{n^\prime,{\bf k}^\prime}|
\psi_{n,{\bf k}}\rangle
=\delta_{n,n^\prime}\, 
\delta^{(D-1)}({\bf k} -{\bf k}^\prime).
\label{4.5}
\end{equation}
On the other hand, in the one-dimensional case the spectrum 
is purely discrete, the normalized eigenfunctions
and eigenvalues being respectively
\begin{equation}
\psi_{n}(x)=\sqrt{-{\kappa\over a_n'}}\, 
{{\rm Ai}(\kappa x+a_n')\over {\rm Ai}(a_n')},
\label{4.6} 
\end{equation}
\begin{equation}
E_{n}=-E_g a_n'\,, \qquad n\in{\mathbb N}.
\label{4.8}
\end{equation}

Let us first analyze in detail the Bose-Einstein 
condensation (BEC) for such a one-dimensional system.
In the grand canonical ensemble the average number of 
particles $N$ at temperature $T$ and chemical potential $\mu$ reads
\begin{equation}
N = \sum_{n=1}^\infty {1 \over 
\exp\left[\beta (E_n-\mu)\right]-1},
\label{4.9a} 
\end{equation}
where, as usual, $\beta=1/k_BT$. The criterion for the 
occurrence of BEC is that the average population of
the excited states remains finite as the chemical potential
approaches the ground state energy from below, i.e., 
\begin{equation}
\lim_{\mu\uparrow E_1}\,N_{\rm ex}=
\lim_{\mu\uparrow E_1}\,\sum_{n=2}^\infty {1\over 
\exp\left[\beta (E_n-\mu)\right]-1}<\infty.
\label{4.9c}
\end{equation}
Notice that the ground state population has been
split off, that being the reason why the above sum begins at $n=2$.
The sequence of eigenvalues (\ref{4.8}) is such that the above 
mentioned BEC criterion is satisfied. Consequently, 
Bose-Einstein condensation is expected to
occur, although, in order to specify the critical 
temperature, it would be
necessary to sum up the series, which, up to our
knowledge, cannot be done analytically.
Nonetheless, one can estimate the 
critical quantities using the 
asymptotic behavior of $E_n$ for large $n$ \cite{AbS}:
\begin{equation}
E_n=-E_g a_n' \sim E_g\left[3\pi(4n-3)/8\right]^{2/3}
,
\qquad n\gg 1.
\label{4.10}
\end{equation}
This corresponds to a density of states of the form
\begin{equation}
\label{rho}
\rho(E)\approx{dn \over dE}
\sim{1 \over \pi}\,E_g^{-3/2}\,E^{1/2},\qquad E\gg E_g\,.
\end{equation}
Since $E_g\propto g^{2/3}$, as $g\to 0$ the energy spectrum becomes
denser and denser and the ground state energy approaches zero. Thus,
in a weak gravitational field 
it is reasonable to
extrapolate in the continuum the density of states (\ref{rho}) down to $E=0$.
We can then approximate the series in Eq.\ (\ref{4.9c})
by an integral, and eventually obtain
\begin{equation}
N_{\rm ex} \sim 
\int_0^\infty {dE \over \pi}\,{E_g^{-3/2}E^{1/2} \over 
\exp\left[\beta(E-\mu)\right]-1}=
4\pi\left(\kappa\lambda_T\right)^{-3} 
g_{3/2}(e^{\beta\mu}),
\label{4.12}
\end{equation}
where $\lambda_T\equiv h/\sqrt{2\pi mk_BT}$ is the
thermal wavelength and
$g_s(x)\equiv\sum_{n=1}^{\infty}n^{-s}\,x^n$ is
the Bose-Einstein function \cite{Hua}. To
obtain the critical temperature, we take the
limit $\mu\to 0$ in Eq.\ (\ref{4.12})
and equate $N_{\rm ex}$ to the total number
of particles in the gas; solving for $T$ then yields
the approximate critical temperature
\begin{equation}
T_{\rm c}\sim {E_g \over k_B}\,(4\pi)^{1/3}
\left({N \over g_{3/2}(1)}\right)^{2/3}.
\label{4.13}
\end{equation}
Below $T_{\rm c}$ the fraction of particles occupying the
ground state is given by
\begin{equation}
{N_0\over N}
=1-{N_{\rm ex} \over N}
=1-\left({T \over T_{\rm c}}\right)^{3/2}.
\label{4.14}
\end{equation}

The reasoning which led us to the conclusion that a one-dimensional
ideal boson gas in a uniform gravitational field displays
BEC can be easily generalized to higher dimensions and
other types of potential. This is the content of the following
theorem.

\begin{theorem}
Suppose the single-particle energy spectrum of an 
ideal boson gas
satisfies the following conditions: (i) there is a gap between
the fundamental and the first excited energy levels, 
i.e., $E_1-E_0=\Delta>0$;
(ii) the single-particle partition function is finite, i.e.,
$Z\equiv\sum_{n=0}^{\infty}d_n\exp(-\beta E_n)<\infty$, $d_n$ being
the finite degeneracy of the $n$-th eigenvalue of the 
single-particle Hamiltonian. 
Then this gas displays Bose-Einstein condensation at finite
temperature.
\end{theorem} 

\noindent
{\it Proof.} If $\mu<E_0$, 
the number of particles in the excited states 
is bounded from above by
\begin{equation}
N_{\rm ex}=\sum_{n=1}^{\infty}\frac{d_n\exp[-\beta(E_n-\mu)]}
{1-\exp[-\beta(E_n-\mu)]}
\le\frac{\exp(\beta\mu)}{1-\exp[-\beta(E_1-\mu)]}
\sum_{n=1}^{\infty}d_n\exp(-\beta E_n).
\end{equation}
Therefore
\begin{equation}
\lim_{\mu\to E_0}\,N_{\rm ex}\le\frac{\exp(\beta E_0)}
{1-\exp(-\beta\Delta)}\left[Z-d_0\exp(-\beta E_0)\right] < \infty ,
\end{equation}
since, by hypothesis, $Z$ and 
$d_0$ are finite and $\Delta > 0$.\hfill{\it Q.E.D.}
\vspace{0.3cm}

We notice that the above statement may 
be generalized to some cases in
which part of the spectrum is continuous 
or there are infinitely
degenerate energy levels. This is done under the
suitable introduction of the density of 
particles in the excited states 
and of the single-particle partition 
function per unit volume. 
Some explicit examples
of this generalization are discussed 
in Ref.\ \cite{GMS2} and in 
Section \ref{Floor} of the present paper.

There are many papers that discuss the problem 
of Bose-Einstein condensation of 
an ideal gas confined in a power-law potential 
\cite{Bagnato2,Li,Yan,Yan2,Salasnich}, mainly 
using some kind of semiclassical 
approximation. In particular, they predict that 
a one-dimensional gas displays 
BEC iff the power-law potential is {\em less} 
confining than the parabolic one, i.e., 
$V(x)\propto x^{\eta}$, $\eta<2$. Theorem 1 shows 
that this condition is too 
strong: BEC occurs for any positive $\eta$. 
It should be clear that the
reason of such a discrepancy is not the 
semiclassical approximation 
{\it per se}, but the substitution of the 
discrete spectrum by a
smooth density of states, which may miss some 
relevant features of the energy spectrum. 

Let us return to the problem of an ideal boson gas
in a uniform gravitational field. We shall now
consider the two- and three-dimensional cases.
Due to the translation invariance along the transverse
direction(s), the proper quantity to be discussed is
the number of particles per unit area
$n^{(D)}\equiv\lim_{L_j\to\infty}N/L_1\cdots L_{D-1}$.
The density of particles in the excited states is then
given by
\begin{eqnarray}
n_{\rm ex}^{(D)} &=& \sum_{j=1}^{\infty}\int
{d^{D-1}k \over 
(2\pi\hbar)^{D-1}}\left\{\exp\left[\beta\left({{\bf k}^2 \over 2m}
-E_g a_j'-\mu\right)\right]-1\right\}^{-1}
\nonumber \\
&=& \lambda_T^{1-D}
\sum_{j=1}^{\infty}g_{(D-1)/2}\left[\exp\beta(E_g a_j'+\mu)
\right],\qquad \mu<-E_g a_1'\,.
\label{nex}
\end{eqnarray}
The integral in Eq.\ (\ref{nex}) is well defined 
for arbitrary $D>1$
due to the condition $\mu<-E_g a_1'$. Now,
since $\lim_{x\to 1}g_s(x)=\infty$ if $s\le 1$,
the first term of the series on the r.h.s.\ of Eq.\ (\ref{nex})
diverges for $D\le 3$ as $\mu\to -E_g a_1'$.
Therefore, a two- or three-dimensional
ideal boson gas in a uniform gravitational field
{\em does not} display Bose-Einstein condensation 
at $T\neq 0$.
 
Some remarks are in order here:

(a) At first sight, Eq.\ (\ref{nex}) seems to imply
absence of BEC in $D=1$ too. It should be noted, however,
that in one dimension there is no integration over 
transverse momenta. Hence, in order to remove the
contribution of the ground state from the sum 
over states in Eq.\ (\ref{nex}), one has to begin it at $j=2$. 
Then $n_{\rm ex}^{(1)}$ $(=N_{\rm ex})$ has a finite 
limit as $\mu\to -E_g a_1'$.

(b) It is easy to see that the absence of BEC in
a two- or three-dimensional ideal boson gas in a
uniform gravitational field in the $x$-direction is due
to the quantization of the motion in that direction. Thus, any
potential $V$ that depends only on $x$, and such that the
one-dimensional Hamiltonian
\begin{equation}
H_x=\frac{p_x^2}{2m}+V(x)
\end{equation}
has a discrete spectrum, will do the job of hindering 
BEC in $D=2,3$.

(c) There are claims in the 
Literature \cite{Hua,Gol,Lam,Groot,Halpern,Gersch,Bagnato1} 
that a three-dimensional ideal boson gas in a uniform field
may undergo BEC at $T\neq 0$. This is an artifact of approximating the sum
in Eq.\ (\ref{nex}) by an integral (remember
that Eq.\ (\ref{nex}) holds true for $D>1$). Indeed, using the density
of states given by Eq.\ (\ref{rho}) we obtain
\begin{eqnarray}
\sum_{j=1}^{\infty}g_{(D-1)/2}\left[\exp\beta(E_ga_j'+\mu)\right]
&\approx&\frac{1}{\pi}\,E_g^{-3/2}\int_0^{\infty}dE\,E^{1/2}
\sum_{n=1}^{\infty}\frac{e^{-n\beta(E-\mu)}}{n^{(D-1)/2}}
\nonumber \\
&=&\frac{1}{\pi}\,(\beta E_g)^{-3/2}\,\Gamma(3/2)
\sum_{n=1}^{\infty}\frac{e^{n\beta\mu}}{n^{(D+2)/2}}
\nonumber \\
&=&4\pi\left(\kappa\lambda_T\right)^{-3}
g_{(D+2)/2}(e^{\beta\mu}).
\label{approxsum}
\end{eqnarray}
Inserting this result into Eq.\ (\ref{nex}), one would be led to the
incorrect conclusion that 
BEC occurs at finite temperature in $D=2$ and $D=3$ in the presence of a 
uniform field, because 
$\lim_{\mu\to 0}\,g_{(D+2)/2}(e^{\beta\mu})<\infty$ if $D>0$.

(d) It should be clear by now that none of our conclusions
so far depends crucially on the use of Neumann boundary condition.
They would remain correct, at least qualitatively, had we used
Dirichlet or Robin boundary condition instead.


\section{$D$-dimensional boson gas interacting with a
point-like impurity at the bottom of the container}
\label{Floor}

In this Section we finally come to the most interesting 
physical case in which, in addition to the gravitational field,
there is a point-like impurity at the bottom of the vessel containing
the gas. As we shall show here,
such an impurity is enough to restore BEC in the
three-dimensional case --- and to allow its existence in the
two-dimensional case, in which it is absent with or without
the gravitational field.
The single-particle Hamiltonian takes now the form
\begin{equation}
H^{(D)}(g,\lambda_D)={{\bf p}^2\over 2m}+mgx
+\lambda_D\,\delta^{(D)}({\bf x})
\equiv H_0^{(D)}(g)+\lambda_D\,\delta^{(D)}({\bf x}).
\label{5.1} 
\end{equation}
Our main task will be to show that the  
$\delta$-potential creates a bound state
in the two- and three-dimensional cases, thus paving the
way for the occurrence of Bose-Einstein
condensation, at variance with
the impurity-free situation discussed in the previous Section. 

Our basic tool to tackle this problem is the Green's function
\begin{equation}
G^{(D)}(z;{\bf x},{\bf x}')=\left<{\bf x}\left|
\left[H^{(D)}(g,\lambda_D)-z\right]^{-1}\right|{\bf x}'\right>,
\qquad z\in\mathbb{C},
\end{equation}
from which it is possible to extract the energy spectrum.
A formal expression for $G^{(D)}(z;{\bf x},{\bf x}')$
can be obtained by solving the Lippmann-Schwinger integral equation,
\begin{equation}
\label{LS}
G^{(D)}(z;{\bf x},{\bf x}')=G_0^{(D)}(z;{\bf x},{\bf x}')-
\int d^Dy\,G_0^{(D)}(z;{\bf x},{\bf y})\,V({\bf y})\,
G^{(D)}(z;{\bf y},{\bf x}'),
\end{equation}
where $G_0^{(D)}$ and $G^{(D)}$ are the Green's functions
associated to $H_0^{(D)}$ and $H^{(D)}=H_0^{(D)}+V({\bf x})$,
respectively. For $V({\bf x})=\lambda_D\,\delta^{(D)}({\bf x})$ 
the integral in Eq.\ (\ref{LS}) can be done trivially, 
resulting in
\begin{equation}
\label{L-S}
G^{(D)}(z;{\bf x},{\bf x}')=G_0^{(D)}(z;{\bf x},{\bf x}')-
\lambda_D\,G_0^{(D)}(z;{\bf x},{\bf 0})\,G^{(D)}(z;{\bf 0},{\bf x}').
\end{equation}
If we now set ${\bf x}={\bf 0}$, we obtain an algebraic
equation for $G^{(D)}(z;{\bf 0},{\bf x}')$. Solving that
equation and inserting the result into Eq.\ (\ref{L-S}),
we end up with
\begin{equation}
\label{Krein}
G^{(D)}(z;{\bf x},{\bf x}')=G_0^{(D)}(z;{\bf x},{\bf x}')-
\frac{G_0^{(D)}(z;{\bf x},{\bf 0})\,G_0^{(D)}(z;{\bf 0},{\bf x}')}
{\frac{1}{\lambda_D}+G_0^{(D)}(z;{\bf 0},{\bf 0})}\,.
\end{equation}
As we shall see
below, $G_0^{(D)}(z;{\bf 0},{\bf 0})$ is formally divergent
for $D\ge 2$, but one can still give a well defined meaning
to Eq.\ (\ref{Krein}) by renormalizing the coupling parameter
$\lambda_D$. The resulting expression, which then makes sense
also for $D=2,3$, is known as
the Krein's formula \cite{Alb} and encodes the one-parameter
family of self-adjoint extensions of the symmetric Hamiltonian operator
$H_0^{(D)}(g)$. This precisely corresponds to the mathematically
rigorous description of the $\delta$-potential.

To complete the construction of $G^{(D)}$ we still have
to obtain the Green's function in the absence
of the impurity. This is done in Appendix \ref{Green},
with the result
\begin{equation}
\label{G0D}
G_0^{(D)}(z;{\bf x},{\bf x}')=-{\pi\kappa \over E_g}
\int\frac{d^{D-1}k}{(2\pi\hbar)^{D-1}}\,
\exp\left\{{i\over\hbar}\,{\bf k}\cdot
({\bf r}^{}-{\bf r}')\right\}
{u[\xi(x_<)]\,v[\xi(x_>)] \over {\rm Ai}'[\xi(0)]}\,,
\end{equation}
where the functions $u(\xi)$ and $v(\xi)$ are defined 
in Eq.\ (\ref{uv}),
$\xi(x)$ is defined in Eq.\ (\ref{xi}), and
$x_<(x_>)={\rm min}({\rm max})\{x,x'\}$. Setting 
${\bf x}={\bf x}'={\bf 0}$ in Eq.\ (\ref{G0D}) we formally obtain
\begin{eqnarray}
G_0^{(D)}(z;{\bf 0},{\bf 0})
&=&-{\kappa\over E_g}\int {d^{D-1}k \over (2\pi\hbar)^{D-1}}\,
{{\rm Ai}\left[\left({\bf k}^2/2mE_g\right)-\left(z/E_g\right)\right]\over
{\rm Ai}'\left[\left({\bf k}^2/2mE_g\right)-\left(z/E_g\right)\right]}
\nonumber \\
&=&-C_D\int_0^\infty dy
\,
y^{(D-3)/2}\, {{\rm Ai}(y-\zeta)\over {\rm Ai}^\prime(y-\zeta)}\,,
\label{5.A}
\end{eqnarray}
where 
\begin{equation}
C_D\equiv{\kappa^D\,(4\pi)^{(1-D)/2}\over E_g\,\Gamma[(D-1)/2]}\,,
\qquad\zeta\equiv \frac{z}{E_g}\,.
\end{equation}
It follows from the asymptotic behavior of the Airy function
${\rm Ai}(x)$ for large $x$ \cite{AbS},
\begin{equation}
\label{asympAiry}
{\rm Ai}(x)\stackrel{x\to\infty}{\sim}\frac{1}{2\sqrt{\pi}x^{1/4}}\,
\exp\left(-\frac{2}{3}\,x^{3/2}\right)\left[1+O(x^{-3/2})\right],
\end{equation}
that the integral in Eq.\ (\ref{5.A}) 
diverges in the UV region for $D\ge 2$, as anticipated.
(The integral is finite in the IR for $D>1$.)

Before we show how to make sense of Eq.\ (\ref{Krein})
for $D=2,3$, let us discuss the one-dimensional case,
which does not need renormalization. 
In this case, the energy spectrum  
can be obtained by solving\footnote{One can easily check 
that the residue of
$G_0^{(1)}(z;x,x')$ at $z=-E_g a_n'$ cancels against the
residue of the second term on the r.h.s.\ of Eq.\ (\ref{Krein})
at the same pole. Therefore, all the poles of $G^{(1)}(z;x,x')$
are given by the solutions to Eq.\ (\ref{pole-1D}).}
\begin{equation}
\label{pole-1D}
\frac{1}{\lambda_1}+G_0^{(1)}(z;0,0)=0,
\end{equation}
or, more explicitly (see Appendix \ref{Green}),
\begin{equation}
\label{K1D}
{1 \over \lambda_1}-{\kappa \over E_g}\,
{{\rm Ai}\left(-z/E_g\right) \over 
{\rm Ai}'\left(-z/E_g\right)}=0.
\label{5.C}
\end{equation} 
This equation is equivalent to 
the imposition of Robin boundary condition at the origin, i.e.,
$\psi'(0)+c\,\psi(0)=0$.
It interpolates between the Neumann boundary condition,
for $\lambda_1\to 0$, and the Dirichlet one, for $\lambda_1\to\infty$.
Any of these boundary conditions
prevents the flow of particles across the origin, so any of them
can be used to represent an impenetrable wall at the bottom of the
container. Nevertheless,
it is more convenient to impose Neumann boundary condition
in the impurity-free case,
because it is then possible to model an
impurity at the bottom of the container by a $\delta$-potential. 
This would not be possible had we imposed Dirichlet 
boundary condition instead. In any case, 
the energy spectrum obtained by solving Eq.\ (\ref{K1D})
will be purely discrete and bounded from below.
As a consequence, we can say that in the one-dimensional case 
the Bose-Einstein condensation
actually occurs at the lowest discrete energy level, although 
the ground state energy itself
as well as the critical quantities are shifted with respect 
to the previously discussed
impurity-free case.

Let us now discuss the two- and three-dimensional cases.
In order to make sense of the denominator in Eq.\ (\ref{Krein}),
we first have to regularize $G_0^{(D)}(z;{\bf 0},{\bf 0})$.
We shall do this by introducing a UV cutoff in Eq.\ (\ref{5.A}), namely,
\begin{equation}
G_{0}^{(D)}(z;{\bf 0},{\bf 0})\to
G_{0}^{(D)}(\Lambda,z;{\bf 0},{\bf 0})
=-C_D\int_0^{\Lambda}dy\,
y^{(D-3)/2}\, {{\rm Ai}(y-\zeta)\over {\rm Ai}^\prime(y-\zeta)}\,.
\end{equation}
We now add to $G_{0}^{(D)}(\Lambda,z;{\bf 0},{\bf 0})$
the integral
\begin{equation}
I_D(\Lambda,z,\alpha)\equiv -C_D\int_0^{\Lambda}dy\,
y^{(D-3)/2}\left(y+\alpha\right)^{-1/2},\qquad\alpha>0.
\end{equation}
It follows from Eq.\ (\ref{asympAiry}) that
\begin{eqnarray}
\frac{{\rm Ai}(y-\zeta)}{{\rm Ai}'(y-\zeta)}&\stackrel{y\to\infty}{\sim}&
-(y-\zeta)^{-1/2}+O\left[(y-\zeta)^{-2}\right]
\nonumber \\
&\sim& -y^{-1/2}+O\left(\zeta y^{-3/2}\right);
\label{asymp2}
\end{eqnarray}
hence, the integrand of 
$G_0^{(D)}(\Lambda,z;{\bf 0},{\bf 0})+I_D(\Lambda,z,\alpha)$
behaves like $y^{(D-6)/2}$ for large $y$. This allows us to
remove the UV regulator (i.e., to take the limit $\Lambda\to\infty$) 
for $D<4$. At the same time, since we have added $I_D$ to $G_0^{(D)}$,
we must subtract it from $\lambda_D^{-1}$ in order to keep
the combination $\lambda_D^{-1}+G_0^{(D)}(z;{\bf 0},{\bf 0})$ 
unaltered. We may then define the renormalized coupling parameter 
$\lambda_D^R$ as
\begin{equation}
\frac{1}{\lambda_D^R}=\lim_{\Lambda\to\infty}\left[
\frac{1}{\lambda_D}-I_D(\Lambda,z,\alpha)\right],
\end{equation}
where it is understood that $\lambda_D$ depends on $\Lambda$
in such a way that the limit exists.
We then finally arrive at a meaningful expression for the
Green's function $G^{(D)}(z;{\bf x},{\bf x}')$ for $D=2,3$,
in which the
denominator of Eq.\ (\ref{Krein}) is replaced by the finite
expression
\begin{equation}
\label{gD}
\texttt{g}_D(\zeta,\alpha,\lambda_D^R)\equiv\frac{1}{\lambda_D^R}
-C_D\int_0^{\infty}dy\,y^{(D-3)/2}
\left[\frac{{\rm Ai}(y-\zeta)}{{\rm Ai}'(y-\zeta)}+
\left(y+\alpha\right)^{-1/2}\right].
\end{equation}

It is possible to show (see Appendix \ref{g=0}) that,
for any finite value of $\lambda_D^R$,
$\texttt{g}_D(\zeta,\alpha,\lambda_D^R)$ has a single zero $\zeta_0$
in the interval $-\infty<\zeta_0<-a_1'$. In physical terms,
this means the existence of a bound state with energy
$E_0=E_g\,\zeta_0$. The rest of the energy spectrum forms a continuum
starting at $E=-E_g a_1'$. The presence of this gap in the
energy spectrum is enough to guarantee the occurrence of BEC.
The proof of this fact is similar to that of Theorem 1,
the only difference being that what saturates in the limit
$\mu\to E_0$ is not $N_{\rm ex}$, but $n_{\rm ex}^{(D)}$.
Some examples of this phenomenon are discussed in detail 
in Ref.\ \cite{GMS2}, where it is also shown how to obtain
the critical quantities. Working in close analogy, one can
obtain an estimate of the critical quantities in the present situation,
taking Eq.\ (\ref{approxsum}) suitably into account. 
If the energy gap created by the impurity is much greater
than the energy splitting due to the gravitational field, 
i.e., $\Delta\equiv-E_g a_1'-E_0\gg -E_g a_2'+E_g a_1'$, one can obtain
a good approximation to the critical temperature $T_{\rm c}$ by
solving the equation
\begin{equation}
\label{Tc}
\lambda_{T_{\rm c}}^{D-1}n^{(D)}=
4\pi\left(\kappa\lambda_{T_{\rm c}}\right)^{-3}
g_{(D+2)/2}\left[\exp(-\Delta/k_BT_{\rm c})\right].
\end{equation}
It is worthwhile to stress that now, because
the bound state energy $E_0$ is strictly below the
continuum threshold $(-E_g a_1')$, we can safely use
Eq.\ (\ref{approxsum}) to estimate the critical quantities in $D=2,3$.

We close this section with a somewhat technical remark.
Aside from being positive, the parameter $\alpha$ in
Eq.\ (\ref{gD}) is arbitrary, and has to be fixed by some 
renormalization prescription. One possibility is the so called
Bergmann-Manuel-Tarrach \cite{BMT} renormalization prescription,
in which the bound state energy $E_0$ 
labels the one-parameter family of self-adjoint extensions
of the symmetric Hamiltonian $H_0^{(D)}(g)$.
Then Eq.\ (\ref{gD}) becomes equivalent to the pair of equations
\begin{equation}
\texttt{g}_D(\zeta,\zeta_0)|_{\rm BMT}
=C_D\int_0^{\infty}dy\,y^{(D-3)/2}\left[\frac{{\rm Ai}(y-\zeta_0)}
{{\rm Ai}'(y-\zeta_0)}-\frac{{\rm Ai}(y-\zeta)}{{\rm Ai}'(y-\zeta)}
\right],
\end{equation}
\begin{equation}
\frac{1}{\lambda_D^R(\alpha)}=C_D\int_0^{\infty}dy\,y^{(D-3)/2}
\left[\frac{{\rm Ai}(y-\zeta_0)}{{\rm Ai}'(y-\zeta_0)}
+(y+\alpha)^{-1/2}\right],
\end{equation}
where $\zeta_0=E_0/E_g<-a_1'$. The parameter $\alpha>0$ is thus
the subtraction point at which the ``running'' coupling parameter
$\lambda_D^R$ is defined.


\section{Conclusions}
\label{Conclusions}

In this paper we have explicitly solved the quantum dynamics 
and studied the thermodynamic equilibrium of an
ideal $D$-dimensional boson gas in the presence 
of a uniform gravitational field
and a point-like impurity at the bottom of the vessel containing
the gas. For convenience, in the present
analysis we have imposed Neumann boundary condition at the bottom
of the container, but our results might be generalized to Dirichlet 
or Robin boundary conditions
without any substantial modification in the physical behavior.
In the impurity-free case it has been shown that Bose-Einstein
condensation at finite temperature is possible only in the 
one-dimensional case and
an estimate of the critical temperature in this case has been 
obtained. It has also been elucidated why the conventional wisdom
that BEC (with a phase separation) might occur in the three-dimensional case
does actually fail: the reason eventually lies in the illegitimate
use of a continuous approximation to the density of states 
in the computation of the average number of particles in the excited states. 

On the other hand, it has been proved that the presence of
a point-like impurity is enough to allow BEC at $T\neq 0$ also in 
two and three dimensions. The reason is that the
impurity creates a
bound state in the single-particle spectrum, where particles 
can now accumulate. It should also be emphasized that a
$\delta$-potential in the presence of a 
uniform field is always attractive in two and three dimensions,
irrespective of the sign of the renormalized coupling parameter. 

The main interest in the study of the present model is in
its exact solvability. Nonetheless, it is evident that
the key physical features here exhibited will persist even
if more realistic impurity potentials are used.
The situation is less clear if one considers an
interacting boson gas (for the general definition of BEC,
applicable to this case, see Ref.\ \cite{Leggett}). 
It is reasonable to assume that
our results still hold if the mean field interaction
between the particles in the gas is smaller than (i) the
energy splitting due to the gravitational field,
and (ii) the energy gap created by the
impurity (if the latter is present). This condition, however,
is likely to be violated as more and more
particles accumulate in the lowest energy level, until
the interaction between the particles cannot be neglected 
anymore. What happens then awaits further investigation.


\acknowledgments

R.M.C.\ acknowledges the kind hospitality of Universit\'a di Bologna
and the financial support from FAPERJ.


\appendix

\section{}
\label{Green}

The Green's function $G_0^{(D)}(z;{\bf x},{\bf x}')$ satisfies
the partial differential equation
\begin{equation}
\label{G}
\left[H_0^{(D)}(g)-z\right]G_0^{(D)}(z;{\bf x},{\bf x}')
=\delta^{(D)}({\bf x}-{\bf x}').
\end{equation}
We can reduce Eq.\ (\ref{G}) to an ordinary differential equation by
Fourier transforming in the transverse coordinates:
\begin{equation}
\label{cG}
\left(-\frac{\hbar^2}{2m}\,\frac{\partial^2}{\partial x^2}
+\frac{{\bf k}^2}{2m}+mgx-z\right)
{\cal G}(z,{\bf k};x,x')=\delta(x-x');
\end{equation}
the Green's function $G_0^{(D)}$ will then be given 
by\footnote{In the one-dimensional case we have instead
$G_0^{(1)}(z;x,x')={\cal G}(z,{\bf k}={\bf 0};x,x')$.}
\begin{equation}
\label{int}
G_0^{(D)}(z;{\bf x},{\bf x}')=\int\frac{d^{D-1}k}
{(2\pi\hbar)^{D-1}}\,
\exp\left\{{i\over\hbar}\,{\bf k}\cdot
({\bf r}-{\bf r}')\right\}
{\cal G}(z,{\bf k};x,x').
\end{equation}
Upon the change of variable 
\begin{equation}
\xi=\kappa x+E_g^{-1}\left({{\bf k}^2 \over 2m}-z\right)
,
\label{xi}
\end{equation}
Eq.\ (\ref{cG}) becomes
\begin{equation}
\label{Airy}
\left(\frac{\partial^2}{\partial\xi^2}-\xi\right)
{\cal G}(\xi,\xi')=-{\kappa\over E_g}\,\delta(\xi-\xi').
\end{equation}

When $\xi\ne\xi'$, Eq.\ (\ref{Airy}) reduces to 
the Airy differential equation. Its solution must satisfy
Neumann boundary condition at $x=0$, i.e.,
$\partial_{\xi}{\cal G}(\xi,\xi')|_{x=0}=0$,
and it must vanish at infinity,
$\lim_{\xi\to\infty}\,{\cal G}(\xi,\xi')=0$.
Thus,
\begin{equation}
\label{sol1}
{\cal G}(\xi,\xi')=C_1\,u(\xi)\,\theta(\xi'-\xi)
+C_2\,v(\xi)\,\theta(\xi-\xi'),
\end{equation}
where 
$\theta(x)$ is the Heaviside step function and
\begin{equation}
u(\xi)\equiv{\rm Bi}'(\xi_0)\,{\rm Ai}(\xi)
-{\rm Ai}'(\xi_0)\,{\rm Bi}(\xi),\qquad
v(\xi)\equiv{\rm Ai}(\xi), 
\label{uv}
\end{equation}
with $\xi_0\equiv\xi(x=0)$.
To fix the constants $C_1$ and $C_2$, one imposes
continuity of ${\cal G}(\xi,\xi')$ at $\xi=\xi'$,
\begin{equation}
\label{cond1}
{\cal G}(\xi'+0,\xi')={\cal G}(\xi'-0,\xi'),
\end{equation}
and a jump in $\partial_{\xi}{\cal G}(\xi,\xi')$ at
the same point, 
\begin{equation}
\label{cond2}
\partial_{\xi}{\cal G}(\xi'+0,\xi')
-\partial_{\xi}{\cal G}(\xi'-0,\xi')=-{\kappa\over E_g},
\end{equation}
obtained by integrating Eq.\ (\ref{Airy})
from $\xi'-\epsilon$ to $\xi'+\epsilon$ and letting
$\epsilon\downarrow 0$.
Applying conditions (\ref{cond1}) and (\ref{cond2})
to the solution (\ref{sol1}), and using the fact that
the Wronskian of $u(\xi)$ and $v(\xi)$ is given by
\begin{equation}
W\{u(\xi),v(\xi)\}=-{\rm Ai}'(\xi_0)\,W\{{\rm Bi}(\xi),{\rm Ai}(\xi)\}
=\frac{1}{\pi}\,{\rm Ai}'(\xi_0),
\end{equation}
we finally obtain
\begin{equation}
\label{cG2}
{\cal G}(\xi,\xi')=-\frac{\pi\kappa\,u(\xi_<)\,v(\xi_>)}
{E_g\,{\rm Ai}'(\xi_0)}\,
,
\end{equation}
where $\xi_<(\xi_>)={\rm min}\,({\rm max})\{\xi,\xi'\}$.
Substituting (\ref{cG2}) into Eq.\ (\ref{int}) gives us the
desired integral representation of Eq.\ (\ref{G0D}) for
$G_0^{(D)}(z;{\bf x},{\bf x}')$.


\section{}
\label{g=0}

Here we show that
$\texttt{g}_D(\zeta,\alpha,\lambda_D^R)$ has one (and only one) zero in the
interval $-\infty<\zeta<-a_1'$. Indeed, for $\zeta$ large
and negative we may use the first line of Eq.\ (\ref{asymp2}) 
to evaluate the integral in Eq.\ (\ref{gD}), obtaining
\begin{equation}
\texttt{g}_2(\zeta,\alpha,\lambda_2^R)\stackrel{\zeta\to-\infty}{\sim}
\frac{1}{\lambda_2^R}-C_2\,\ln\left(-\frac{\zeta}{\alpha}\right),
\end{equation}
\begin{equation}
\texttt{g}_3(\zeta,\alpha,\lambda_3^R)\stackrel{\zeta\to-\infty}{\sim}
\frac{1}{\lambda_3^R}-2\,C_3\left(\sqrt{-\zeta}-\sqrt{\alpha}\right).
\end{equation}
In both cases, 
$\lim_{\zeta\to-\infty}\texttt{g}_D(\zeta,\alpha,\lambda_D^R)=-\infty$.
On the other hand, the integral in Eq.\ (\ref{gD}) 
becomes divergent at the origin for $D\le 3$ if $\zeta\uparrow-a_1'$, as
\begin{equation}
\frac{{\rm Ai}(y+a_1')}{{\rm Ai}'(y+a_1')}\stackrel{y\to 0}{\sim}
\frac{{\rm Ai}(a_1')}{{\rm Ai}''(a_1')\,y}
=\frac{1}{a_1' y}\,.
\end{equation}
(The last equality is a consequence of Airy differential
equation.) Since $a_1'<0$, it follows that
$\lim_{\zeta\uparrow-a_1'}\texttt{g}_D(\zeta,\alpha,\lambda_D^R)=+\infty$
($D=2,3$). By continuity, we may conclude that
$\texttt{g}_D(\zeta,\alpha,\lambda_D^R)$ vanishes at least once
in the interval $-\infty<\zeta<-a_1'$. To show that
it vanishes only once, it suffices to prove that 
$\texttt{g}_D(\zeta,\alpha,\lambda_D^R)$
is a monotonically increasing function of $\zeta$ in that interval.
This follows from the identity
\begin{eqnarray}
\frac{\partial}{\partial\zeta}\,\texttt{g}_D(\zeta,\alpha,\lambda_D^R)
&=&E_g\,\frac{\partial}{\partial z}\left[\frac{1}{\lambda_D}
+G_0^{(D)}(z;{\bf 0},{\bf 0})\right]
\nonumber \\
&=&E_g\left<{\bf 0}\left|\left[H_0^{(D)}(g)-z\right]^{-2}\right|{\bf 0}
\right>.
\end{eqnarray}
It shows that $\partial_{\zeta}\texttt{g}_D(\zeta,\alpha,\lambda_D^R)>0$
if $z$ is real and does not belong to the spectrum of
$H_0^{(D)}(g)$. This occurs, as we have seen in Section \ref{no-imp},
for $z<-E_g a_1'$, or $\zeta<-a_1'$.



\begin{references}

\bibitem{Hua} K. Huang, {\it Statistical Mechanics} 
(Wiley, New York, 1987).

\bibitem{Pat} R. K. Pathria, {\it Statistical Mechanics} 
(Pergamon Press, Oxford, 1972).

\bibitem{IGH} L. C. Ioriatti, Jr., S. Goulart Rosa, Jr. and O. Hip\'olito, Am. J. Phys. {\bf 44}, 744 (1976).


\bibitem{GMS2} P. Giacconi, F. Maltoni and R. Soldati, 
Phys. Lett. A {\bf 279}, 12 (2001).

\bibitem{Gol} L. Goldstein, J. Chem. Phys. {\bf 9}, 273 (1941).

\bibitem{Lam} W. Lamb and A. Nordsieck, 
Phys. Rev. {\bf 59}, 677 (1941).

\bibitem{Groot} S. R. de Groot, G. J. Hooyman and
C. A. ten Seldam, Proc. R. Soc. London A {\bf 203},
266 (1950).

\bibitem{Halpern} O. Halpern, Phys. Rev. {\bf 86}, 126 (1952);
{\bf 87}, 520 (1952).

\bibitem{Gersch} H. A. Gersch, J. Chem. Phys. {\bf 27},
928 (1957). 

\bibitem{Bagnato1} V. Bagnato, D. E. Pritchard and
D. Kleppner, Phys. Rev. A {\bf 35}, 4354 (1987).

\bibitem{Alb} S. Albeverio, F. Gesztesy, R. H{\o}egh-Krohn
and H. Holden, {\it Solvable Models in Quantum Mechanics}
(Springer-Verlag, New York, 1988) pp. 109--110, 357--358.

\bibitem{Jac} R. Jackiw, in {\it M. A. B. B\'eg Memorial Volume},
edited by A. Ali and P. Hoodbhoy (World Scientific, Singapore, 1991).

\bibitem{ReS} M. Reed and B. Simon, 
{\it Methods of Modern Mathematical Physics}
(Academic Press, Orlando, 1987), Vol. 2.

\bibitem{Bagnato2} V. Bagnato and D. Kleppner,
Phys. Rev. A {\bf 44}, 7439 (1991).

\bibitem{Li} M. Li, L. Chen and C. Chen,
Phys. Rev. A {\bf 59}, 3109 (1999).

\bibitem{Yan} Z. Yan, Phys. Rev. A {\bf 59}, 4657 (1999).

\bibitem{Yan2} Z. Yan, M. Li, L. Chen, C. Chen and J. Chen, 
J. Phys. A {\bf 32}, 4069 (1999).

\bibitem{Salasnich} L. Salasnich, 
J. Math. Phys. {\bf 41}, 8016 (2000).

\bibitem{AbS} {\it Handbook of Mathematical Functions},
edited by M. Abramowitz and I. A. Stegun
(Dover, New York, 1972) pp. 446--452.

\bibitem{BMT} O. Bergmann, Phys. Rev. D {\bf 46}, 5474 (1992);
C. Manuel and R. Tarrach, Phys. Lett. B {\bf 268}, 222 (1991).

\bibitem{Leggett} A. J. Leggett,
Rev. Mod. Phys. {\bf 73}, 307 (2001).

\end{references}
\end{document}